\documentclass[10pt,journal]{IEEEtran}

\ifCLASSOPTIONcompsoc
  \usepackage[nocompress]{cite}
\else
  \usepackage{cite}
\fi
\ifCLASSINFOpdf
\else
\fi
\usepackage{amsmath}
\usepackage{dsfont}
\usepackage{tikz}
\usepackage[T1]{fontenc}
\usepackage[utf8]{inputenc}
\usepackage{pgfplots}
\usepackage{grffile}
\usepackage{bm}
\pgfplotsset{compat=newest}
\usetikzlibrary{plotmarks}
\usetikzlibrary{arrows.meta}
\usepgfplotslibrary{patchplots}
\usepackage{algorithm}
\usepackage{algorithmic}
\floatname{algorithm}{Procedure}

\usepackage{makecell}

\begin{document}
%
\title{Transceive Phase Corrected Contrast Source Inversion-Electrical Properties Tomography}
%
%
%
%

\author{Peter~R.S.~Stijnman,~
        Stefano~Mandija,~
        Patrick S. Fuchs,~
        Cornelis~A.T.~van~den~Berg,~
        and~Rob~F.~Remis~
\IEEEcompsocitemizethanks{\IEEEcompsocthanksitem $^*$P.R.S. Stijnman is with Computational Imaging Group for MRI diagnostics and therapy, Centre for Image Sciences, University Medical Center Utrecht and the Department of Biomedical Engineering, Eindhoven University of Technology, Eindhoven. (email: P.R.S.Stijnman@umcutrecht.nl)\protect
\IEEEcompsocthanksitem S. Mandija and C.A.T. van den Berg are with the Computational Imaging Group for MRI diagnostics and therapy, Centre for Image Sciences, University Medical Center Utrecht.
\IEEEcompsocthanksitem P.S. Fuchs and R.F. Remis are with the Circuit \& Systems Group of the Electrical Engineering, Mathematics and Computer Science Faculty of the Delft University of Technology.}
}

%
%

\markboth{\empty}%
{Stijnman \MakeLowercase{\textit{et al.}}: Transceive Phase Corrected Contrast Source Inversion-Electrical Properties Tomography}
%



\IEEEtitleabstractindextext{%
\begin{abstract}
Magnetic resonance imaging (MRI) based electrical properties tomography (EPT) is the quantification of the conductivity and permittivity of different tissues. These electrical properties can be obtained through different reconstruction methods and can be used as a contrast mechanism. The work presented here continues from the two-dimensional CSI-EPT algorithm which was shown to work with two-dimensional Matlab based simulations. The existing CSI-EPT algorithm is reformulated to use the transceive phase rather than relying on the transceive phase assumption. This is achieved by implementing a forward problem, computing the receive phase, into the inverse minimization problem, i.e. retrieving the electrical properties. Furthermore, the radio frequency (RF) shield is numerically implemented to model the RF fields inside the MRI more accurately. Afterwards, the algorithm is tested with three-dimensional FDTD simulations to investigate if the two-dimensional CSI-EPT can retrieve the electrical properties for three-dimensional RF fields. Finally, an MR experiment with a  phantom is performed to show the potential for this method. From the results of the two-dimensional Matlab simulations it is seen that CSI-EPT can reconstruct the electrical properties using MRI accessible quantities. In the three-dimensional simulations it is observed that the electrical properties are underestimated, nonetheless, CSI-EPT is more precise than the standard Helmholtz based methods. Finally, the first CSI-EPT results using measured data are shown. The results for the reconstruction using measured data were of the same quality as the results from the FDTD simulation.

\end{abstract}

}

\maketitle

\IEEEdisplaynontitleabstractindextext

%
\IEEEpeerreviewmaketitle

\ifCLASSOPTIONcompsoc
\IEEEraisesectionheading{\section{Introduction}\label{sec:introduction}}
\else
\section{Introduction}
\label{sec:introduction}
\fi

%
%
%
%

\IEEEPARstart{E}{lectrical} properties tomography (EPT) is the technique of imaging the conductivity and permittivity of tissues. This is achieved in a noninvasive manner through MRI-based mapping of the circularly polarized magnetic component ($B_1^+$, the transmit efficiency) of the transmit radio frequency (RF) field. The acquired conductivity and permittivity can be used as a contrast mechanism, especially the conductivity can be used as an endogenous biomarker in oncology\cite{biomarker}. Further, the conductivity and permittivity both are important in computing the specific absorption rate (SAR), which defines the amount of deposited energy during the MRI exam and relates directly to the heating of the tissue under examination \cite{PennesSAR}.

There are a variety of different EPT approaches that have been published. A large group of these approaches are derivative-based that stem from the Helmholtz equation for magnetic fields \cite{Katscher2009,moreEPT,gEPT}. In these approaches a second order derivative using finite difference kernels is applied on the measured $B_1^+$ fields. Furthermore, piece-wise constant electrical properties are assumed which together with finite difference kernels introduce errors in the reconstruction of the electrical properties most notably at tissue boundaries \cite{Smandija17}.

Next to these derivative approaches, there are also integral approaches proposed \cite{Balidemaj2015,VIE} where the objective is to minimize a cost function by iteratively updating the electrical properties. Compared to derivative approaches, integral approaches are more robust with respect to noisy input data and heterogeneous electrical property maps, i.e. tissue boundaries, are easily incorporated. However, these benefits come at the expense of a higher computational cost and generally integral-based optimization methods are more difficult to implement.

We take the contrast source inversion (CSI) method \cite{Abubaker2001} as a starting point, which was formulated for EPT in MRI quantities in \cite{Balidemaj2015}. CSI-EPT takes a three step approach to retrieve the electrical properties. The first step computes a weighted electric field, the contrast source, from the measured $B_1^+$ data. The second step computes the corresponding electric field using the contrast source. The last step combines the contrast source and the electric field in order to retrieve the contrast, i.e. the electrical properties. The three steps are used in an iterative fashion to minimize a cost functional. To compute the separate steps of this method integral operators are used on the contrast source. The measured input data, $B_1^+$, is defined as 
\begin{equation}
    B_1^+ = \frac{B_x + jB_y}{2},
\end{equation}
where $B_x$ and $B_y$ are the $x$- and $y$-components of the magnetic flux density respectively, and $j$ is the imaginary unit \cite{haacke1999magnetic}. The $B_1^+$ field is one of the inputs for the CSI-EPT algorithm that is required to retrieve the electrical properties. The other inputs that are necessary are the incident electric and magnetic field. These are the RF fields of the empty transmit coil, i.e. when the object/patient is not present. These RF fields can be obtained through electromagnetic simulations and will vary from coil to coil.

Similar to other EPT reconstruction methods \cite{Katscher2009,gEPT,VIE}, the complex B1+ field is required in CSI-EPT. CSI-EPT is not the only method that requires complex $B_1^+$ data, most of the EPT methods that are published use the complex $B_1^+$ data as input. The complex $B_1^+$ entails that the magnitude and the phase of the $B_1^+$ field need to be measured. The magnitude can be measured using a multitude of different MRI pulse sequences \cite{Yarnykh2007,Nehrke2012}. The $B_1^+$ phase, however, cannot be directly measured with MRI as the superposition of the $B_1^+$ phase and $B_1^-$ phase, the receive sensitivity, is inherently measured \cite{VanLier2012}. The combination of the $B_1^+$ and $B_1^-$ phase is called the transceive phase. To obtain the $B_1^+$ phase from the transceive phase an assumption is used called the transceive phase assumption (TPA). This assumption states that the quadrature $B_1^+$ transmit and the reverse quadrature $B_1^-$ receive phase of the birdcage coil are equal, therefore, dividing the MR measured transceive phase by a factor of 2 results in the required transmit phase.

As with any assumption there are conditions that need to be met before the TPA can properly be used. For the TPA to be valid the scattered $B_1^+$ fields created by the two ports in a birdcage coil should be equal to each other in order to cancel. Or the contribution of the scattered $B_1^+$ fields compared to the incident fields should be very small. If one of these conditions is met the contribution of the scattered fields to the $B_1^+$ phase can be neglected. An example when these conditions are met is when simple symmetrical geometries, e.g. homogeneous cylinders, load the transmit coil. For more complex structures, non-symmetrical objects or multiple tissue interfaces, these conditions are not met and the assumption is not valid \cite{VanLier2014}. This occurs especially at higher static magnetic field strengths due to the larger magnitudes of scattered RF fields. Therefore, the influence of the scattered RF fields on the $B_1^+$ phase cannot be neglected anymore.

There are methods published that do not use the TPA. One of these methods is the gradient EPT method \cite{Calcagno2016}, where a transmit array is used and the $B_1^+$ phases per transmit channel are taken relative to a reference channel. For this method to work seed points, known points inside the contrast, are needed. Further, in \cite{katscherlocalSAR} the TPA is not used, however, there a comparable assumption is made; the scattered RF fields should be negligible. For CSI-EPT it was previously shown that using the transceive phase rather than the $B_1^+$ phase showed almost no degradation in the quality of the reconstructed electrical properties. However, in contrary to what was shown in \cite{Balidemaj2015}, the transceive phase does have an effect on the CSI-EPT reconstruction and will be investigated here. 

When observing the remaining inputs of the CSI-EPT algorithm, the incident electric and magnetic fields, we should notice how these were acquired previously. Both of the incident fields were modeled using two-dimensional line source simulations. However, the CSI-EPT algorithm as described in \cite{Balidemaj2015} has no RF-shield included in the model. The RF-shield affects the scaling of the incident fields considerably and is therefore required to scale the fields correctly with respect to the current running through the line sources. This scaling is especially relevant for the practical implementation of CSI-EPT, when the incident fields are not known and need to be modelled.

A solution to implement the RF-shield into the CSI-EPT scheme has been proposed in \cite{arduino17}, where instead of the free space Green's tensor functions the Green's tensor functions in the the presence of a circular perfect electrical conductor (PEC), which represents the RF-shield has been used. This solution showed great improvement, however, the downside of using this method is that the computation time of the CSI-EPT algorithm will increase drastically. 

The effect of the RF-shield on the scattered field created by the subject in the MR is minimal, the magnitude of the scattered field decreases rapidly outside of the subject. Therefore, a numerical method to implement the RF-shield in the incident fields used as input could be sufficient. Ans since integral methods are known for having longer computation times than their derivative counterparts we will explore a numerical approximation to implement the RF-shield in this work. Such an approximation allows the use of the free space Green's function, and will essentially not increase the computation time. Further, in this work we will investigate what effect the transceive phase has on the reconstruction of the electrical properties. To solve the fact that the $B_1^+$ phase is not measurable, we reformulate the CSI-EPT algorithm to use the $B_1^+$ magnitude and transceive phase as input rather than the complex $B_1^+$ data. This will eliminate the need for the TPA and instead directly use the measured data from the MRI resulting in a more practically usable methodology. As a result will we present the first reconstructions of electrical properties with MRI acquired data using CSI-EPT in this work.

\section{Theory}
\subsection{Transceive Phase Correction}
In this section we reformulate the CSI-EPT algorithm as described in \cite{Balidemaj2015}. Starting with defining the contrast source as
\begin{equation}
\label{eq:contrastSource}
    w(\bm{\rho}) = \chi (\bm{\rho}) E_z(\bm{\rho}),
\end{equation}
where $E_z$ is the $z$-component of the total electric field, and $\chi$ is the contrast with respect to free space which is defined as 
\begin{equation}
\label{eq:contrast}
    \chi (\bm{\rho})=\varepsilon_r(\bm{\rho}) -1 + \frac{\sigma(\bm{\rho})}{j\omega \varepsilon_0},
\end{equation}
with $\bm{\rho}$ as the position vector, $\varepsilon_0$ and $\varepsilon_r$ as the permittivity in vacuum and relative permittivity respectively, $\sigma$ is the conductivity and $\omega$ is the angular Larmor frequency. Reconstructing the contrast is the goal of this method since the electrical properties can directly be calculated once the contrast is known.
The contrast and contrast source are the two parameters that are iteratively updated in the former CSI-EPT scheme. This is realized by minimizing the cost functional presented in \cite{Balidemaj2015} given by
\begin{equation}
\begin{split}
    \label{eq:oldcostfunction}
    F(w_l,\chi) &= \eta_S \sum\limits_l || B_{1,l}^+ - (B_{1,l}^{+,\text{inc}} + G_S^+\{w_l\})||_S^2 \\
    &+ \eta_D \sum\limits_l||\chi ( E_{z,l}^{\text{inc}} + G_D\{w_l\}) - w_l ||_D^2,
    \end{split}
\end{equation}
where $l \geq 1$ indicates various transmit channels including linear combinations such as a standard quadrature drive, $\eta_{S,D}$ are normalization factors for the data functional (first term on the right-hand side of (\ref{eq:oldcostfunction})) and object functional (second term on the right-hand side of (\ref{eq:oldcostfunction})). It should be noted that for MRI the domain where the object is located, $D$, is the same as the location where the data is collected, $S$. Further, $G_S^+$ and $G_D$ are integral operators that map the contrast source to the scattered $B_1^+$ field, $B_1^{+,sc}$, and the scattered electric field, $E_z^{sc}$, respectively. These integral operators are given by
\begin{subequations}
    \begin{align}
        \begin{split}
        E_z^{\text{sc}}(\bm{\rho})   &= G_D\{w(\bm{\rho'})\}\\
        &= k_0^2 \int_{\bm{\rho} \in \mathds{D}} \hat{G}(\bm{\rho} - \bm{\rho'})  w(\bm{\rho'}) dV, \end{split}\\
        \begin{split}
        B_1^{+,\text{sc}}(\bm{\rho}) &= G_S^+\{w(\bm{\rho'})\} \\
        &= \frac{\omega}{2c_0^2}(\partial_x + j\partial_y) \int_{\bm{\rho} \in \mathds{D}} \hat{G}(\bm{\rho} - \bm{\rho'})  w(\bm{\rho'}) dV.
        \end{split}
    \end{align}
\end{subequations}
Here, $k_0$ and $c_0$ are the wavenumber and the speed of light in vacuum, respectively. The spatial derivatives with respect to $x$ and $y$ are indicated with $\partial_{x,y}$. The source locations are defined by $\bm{\rho'}$ and $\hat{G}(\bm{\rho} - \bm{\rho'})$ is defined as the two-dimensional free space Green's function, which is given by
\begin{equation}
    \label{eq:2DGreens}
    \hat{G}(\bm{\rho} - \bm{\rho'}) = - \frac{j}{4} H_0^{(2)}(k_0 |(\bm{\rho} - \bm{\rho'})|),
\end{equation}
with, finally, $H_0^{(2)}$ defining the zeroth order Hankel function of the second kind.

The problem with the cost functional given by (\ref{eq:oldcostfunction}) is that the phase of the $B_1^+$ field is not directly accessible through measurements. What can be measured is the so-called transceive phase $\phi_\pm$. This transceive phase consists of the transmit phase $\phi_+$ and receive phase $\phi_-$ according to
\begin{equation}
    \phi_\pm = \phi_+ +\phi_-.
\end{equation}
Therefore, we can write the $B_1^+$ field in its polar form as
\begin{equation}
    \label{eq:B1Polar}
    \begin{split}
    B_1^+ &= |B_1^+|e^{j\phi_+}    \\
    &= |B_1^+|e^{j\phi_{\pm}}e^{-j\phi_-}.
    \end{split}
\end{equation}
Substituting (\ref{eq:B1Polar}) into (\ref{eq:oldcostfunction}) results in a new cost functional given by
\begin{equation}
\label{eq:costfunctionnew}
\begin{split}
    &F(w_l,\chi,\phi_-)= \\
    &\eta_S \sum\limits_l || |B_{1,l}^{+}|e^{j\phi_{\pm}}e^{-j\phi_-} - (B_1^{+,\text{inc}} + G_S^+\{w_l\})||_S^2\\
    &+ \eta_D \sum\limits_l||\chi ( E_{z,l}^{\text{inc}} + G_D\{w_l\}) - w_l ||_D^2.\\
\end{split}
\end{equation}
The transmit phase is written in terms of the transceive (known) and receive phase (unknown). Iterating through the algorithm now proceeds in a similar fashion as standard CSI-EPT, except that the receive phase is iteratively updated using the estimates of the contrast function and electric field strength at the current iteration, $n$. 

In this section, the electric and magnetic RF fields are the fields as encountered in the receive state of the birdcage coil. During transmission the birdcage coil is fed in quadrature mode (i.e. a $90^{\circ}$ phase difference between the two ports) and this creates a circularly polarized $B_1$ field that is efficient in tipping the spins. When the desired signal is created the birdcage coil is switched to a receive state which means the $90^{\circ}$ phase difference becomes $-90^{\circ}$ phase difference. This results in a circularly polarized field that is efficient in receiving the signal. This receiving state of the birdcage coil is called reverse quadrature or anti-quadrature. This mode of the transmit coil creates different RF fields. The incident field display a simple phase shift for a birdcage coil setup compared with forward quadrature, but the scattered fields are inherently different which is the origin as to why the TPA is not valid when these scattered fields are large in magnitude. 

To compute the scattered receive field at each iteration we use
\begin{equation}
\label{eq:b1-re}
    \begin{split}
    &B_{1,n}^{\ddagger}(\bm{\rho})) = G_S^-\{\chi_{n-1}(\bm{\rho'}) E_{z,n}(\bm{\rho'}) \}\\
    = -\frac{\omega}{2c_0^2}&(\partial_x - j\partial_y) \int_{\bm{\rho} \in \mathds{D}} \hat{G}(\bm{\rho} - \bm{\rho'})  \chi_{n-1}(\bm{\rho'}) E_{z,n}(\bm{\rho'}) dV,
    \end{split}
\end{equation}
where $B_{1,n}^{-,\ddagger}$ defines the complex conjugate of the scattered receive field and $E_{z,n}$, the electric field during reception at iteration $n$, is defined as
\begin{equation}
\label{eq:ezre}
    E_{z,n}(\bm{\rho}) = G_D\{\chi_{n-1}(\bm{\rho}) E_{z,n-1}(\bm{\rho}) \} + E_z^{\text{inc}}(\bm{\rho}). 
\end{equation}
Here $E_z^{\text{inc}}$, the incident electric field during reception, is modeled together with $B_1^{-,\text{inc}}$ for an empty coil, e.g. in reverse quadrature mode for a birdcage coil. Since the incident fields for the transmit state are already modelled for each port the incident receive fields are acquired by driving these ports in reverse quadrature.

By adding the receive phase as an extra unknown into the minimization problem, the minimization becomes more difficult and the computation time is increased since the number of required integral operations is increased to obtain the receive phase. However, the TPA is no longer required to reconstruct the contrast from the $B_1^+$ data.
The new pseudoalgorithm is shown below with $\dagger$ indicating the quantities during the receive state of the birdcage coil.

\begin{algorithm}
\label{alg:alg}
\caption{Transceive phase corrected CSI-EPT}
\begin{algorithmic}
\REQUIRE Compute $w_0$, $w_0^{\dagger}$, $\chi_0$ and the incident RF-fields\\
\FOR{$n=1$ to maxIterations}
    \STATE \textbf{Step 1} Update the contrast sources:\\
        \STATE Compute $g_n^{w}$ for $F_n^{R}(w_{n-1},\chi_{n-1},\varphi_{n-1})$\\
        \STATE Compute the Polak-Ribière update direction\\
        \STATE Compute the stepsize\\
        \STATE Update the contrast source\\
    \STATE \textbf{Step 2} Update $E_z$\\
    \STATE \textbf{Step 3 }Update the contrast:\\
        \STATE Compute $g_n^{\chi}$ for $F_n^{R}(w_{n},\chi_{n-1},\varphi_{n-1})$\\
        \STATE Compute the Polak-Ribière update direction\\
        \STATE Compute the stepsize\\
        \STATE Update the contrast\\
    \STATE \textbf{Step 4} Update the receive phase:\\
        \STATE Solve for $E_z^{\dagger}$ with $\chi_n$\\
        \STATE Compute $B_1^{-,\dagger}$\\
        \STATE Extract the receive phase\\
    \STATE \textbf{Step 5} Update the residuals and normalization factors\\
\IF{$F_n \leq tolerance$}
    \STATE Break\\
\ENDIF    
\ENDFOR
\ENSURE $\chi_n$, $w_n$, $B_1^{+}$, $B_1^{-,\dagger}$, $E_z$ and $E_z^{\dagger}$\\
\end{algorithmic}
\end{algorithm}

\subsection{Numerical Implementation Of The RF-Shield}

Inside every MRI system there is an RF-shield present. The purpose of the RF-shield is to limit the interaction between external RF signals and the MRI acquisition. The RF-shield changes the RF-fields produced by the transmit coil and as previously stated, the incident electric and magnetic RF-fields are required as input for the CSI-EPT algorithm. Therefore, including the RF-shield in the model that simulates these incident RF-fields is required. The copper RF-shield can be approximated in simulations as a perfect electrically conducting material (PEC).  

In \cite{arduino17} the RF-shield was included into the Green's tensor functions. This is an accurate way to do this, however, different MRI systems would require different Green's tensor functions. Furthermore, the spatial invariance of the free space Green's tensor function is lost, $\hat{G}(\rho - \rho') \rightarrow \hat{G}(\rho,\rho')$ and this results in a significant increase in computation time.

\begin{figure}
\definecolor{mycolor1}{rgb}{0.00000,0.44700,0.74100}%
\definecolor{mycolor2}{rgb}{0.85000,0.32500,0.09800}%
\begin{tikzpicture}

\begin{axis}[%
ticks=none,
width=1.4in,
height=1.4in,
at={(1in,3.892in)},
scale only axis,
point meta min=0,
point meta max=1,
xmin=0.0983,
xmax=0.8023,
xtick={0.2,0.4,0.6,0.8},
xticklabels={\empty},
ymin=0.0983,
ymax=0.8023,
ytick={0.1,0.2,0.3,0.4,0.5,0.6,0.7,0.8},
yticklabels={\empty},
axis background/.style={fill=white},
title style={font=\bfseries},
title={Line source locations}
]
\addplot [color=mycolor1, draw=none, mark=x, mark options={solid, mycolor1}, forget plot]
  table[row sep=crcr]{%
0.4503	0.0983\\
0.585004568192512	0.125094404556027\\
0.699201586977665	0.201398413022335\\
0.775505595443973	0.315595431807488\\
0.8023	0.4503\\
0.775505595443973	0.585004568192512\\
0.699201586977665	0.699201586977665\\
0.585004568192512	0.775505595443973\\
0.4503	0.8023\\
0.315595431807488	0.775505595443973\\
0.201398413022335	0.699201586977665\\
0.125094404556027	0.585004568192512\\
0.0983	0.4503\\
0.125094404556027	0.315595431807488\\
0.201398413022335	0.201398413022335\\
0.315595431807488	0.125094404556027\\
};
\end{axis}

\begin{axis}[%
ticks=none,
width=1.4in,
height=1.4in,
at={(1in,1.533in)},
scale only axis,
point meta min=0,
point meta max=1,
xmin=0.0582197443181818,
xmax=0.842380255681818,
xtick={0.2,0.4,0.6,0.8},
xticklabels={\empty},
ymin=0.0582197443181818,
ymax=0.842380255681818,
ytick={0.1,0.2,0.3,0.4,0.5,0.6,0.7,0.8},
yticklabels={\empty},
axis background/.style={fill=white},
title style={font=\bfseries, align=center},
title={Line and mirror\\source locations},
legend style={at={(0.15,1.35)}, anchor=south west, legend cell align=left, align=left, draw=white!15!black}
]
\addlegendimage{only marks, mark=x,color=mycolor1}
\addlegendimage{only marks, mark=x,color=mycolor2}
\addplot [color=mycolor1, draw=none, mark=x, mark options={solid, mycolor1}]
  table[row sep=crcr]{%
0.4503	0.0983\\
0.585004568192512	0.125094404556027\\
0.699201586977665	0.201398413022335\\
0.775505595443973	0.315595431807488\\
0.8023	0.4503\\
0.775505595443973	0.585004568192512\\
0.699201586977665	0.699201586977665\\
0.585004568192512	0.775505595443973\\
0.4503	0.8023\\
0.315595431807488	0.775505595443973\\
0.201398413022335	0.699201586977665\\
0.125094404556027	0.585004568192512\\
0.0983	0.4503\\
0.125094404556027	0.315595431807488\\
0.201398413022335	0.201398413022335\\
0.315595431807488	0.125094404556027\\
};
\addlegendentry{Line sources}

\addplot [color=mycolor2, draw=none, mark=x, mark options={solid, mycolor2}]
  table[row sep=crcr]{%
0.4503	0.0582197443181818\\
0.6003426180069	0.088065076673776\\
0.727542607561969	0.173057392438031\\
0.812534923326224	0.3002573819931\\
0.842380255681818	0.4503\\
0.812534923326224	0.6003426180069\\
0.727542607561969	0.727542607561969\\
0.6003426180069	0.812534923326224\\
0.4503	0.842380255681818\\
0.3002573819931	0.812534923326224\\
0.173057392438031	0.727542607561969\\
0.0880650766737761	0.6003426180069\\
0.0582197443181818	0.4503\\
0.0880650766737761	0.3002573819931\\
0.173057392438031	0.173057392438031\\
0.3002573819931	0.0880650766737761\\
};
\addlegendentry{Mirror sources}

\end{axis}

\begin{axis}[%
ticks=none,
width=1.4in,
height=1.4in,
at={(2.6in,3.892in)},
scale only axis,
point meta min=0,
point meta max=1,
axis on top,
xmin=0.5,
xmax=474.5,
xtick={100,200,300,400},
xticklabels={\empty},
y dir=reverse,
ymin=0.5,
ymax=474.5,
ytick={100,200,300,400},
yticklabels={\empty},
axis background/.style={fill=white},
title style={font=\bfseries},
title={$\text{$|$E}_\text{z}^{\text{inc}}\text{$|$}$},
colormap={mymap}{[1pt] rgb(0pt)=(0.2081,0.1663,0.5292); rgb(1pt)=(0.211624,0.189781,0.577676); rgb(2pt)=(0.212252,0.213771,0.626971); rgb(3pt)=(0.2081,0.2386,0.677086); rgb(4pt)=(0.195905,0.264457,0.7279); rgb(5pt)=(0.170729,0.291938,0.779248); rgb(6pt)=(0.125271,0.324243,0.830271); rgb(7pt)=(0.0591333,0.359833,0.868333); rgb(8pt)=(0.0116952,0.38751,0.881957); rgb(9pt)=(0.00595714,0.408614,0.882843); rgb(10pt)=(0.0165143,0.4266,0.878633); rgb(11pt)=(0.0328524,0.443043,0.871957); rgb(12pt)=(0.0498143,0.458571,0.864057); rgb(13pt)=(0.0629333,0.47369,0.855438); rgb(14pt)=(0.0722667,0.488667,0.8467); rgb(15pt)=(0.0779429,0.503986,0.838371); rgb(16pt)=(0.0793476,0.520024,0.831181); rgb(17pt)=(0.0749429,0.537543,0.826271); rgb(18pt)=(0.0640571,0.556986,0.823957); rgb(19pt)=(0.0487714,0.577224,0.822829); rgb(20pt)=(0.0343429,0.596581,0.819852); rgb(21pt)=(0.0265,0.6137,0.8135); rgb(22pt)=(0.0238905,0.628662,0.803762); rgb(23pt)=(0.0230905,0.641786,0.791267); rgb(24pt)=(0.0227714,0.653486,0.776757); rgb(25pt)=(0.0266619,0.664195,0.760719); rgb(26pt)=(0.0383714,0.674271,0.743552); rgb(27pt)=(0.0589714,0.683757,0.725386); rgb(28pt)=(0.0843,0.692833,0.706167); rgb(29pt)=(0.113295,0.7015,0.685857); rgb(30pt)=(0.145271,0.709757,0.664629); rgb(31pt)=(0.180133,0.717657,0.642433); rgb(32pt)=(0.217829,0.725043,0.619262); rgb(33pt)=(0.258643,0.731714,0.595429); rgb(34pt)=(0.302171,0.737605,0.571186); rgb(35pt)=(0.348167,0.742433,0.547267); rgb(36pt)=(0.395257,0.7459,0.524443); rgb(37pt)=(0.44201,0.748081,0.503314); rgb(38pt)=(0.487124,0.749062,0.483976); rgb(39pt)=(0.530029,0.749114,0.466114); rgb(40pt)=(0.570857,0.748519,0.44939); rgb(41pt)=(0.609852,0.747314,0.433686); rgb(42pt)=(0.6473,0.7456,0.4188); rgb(43pt)=(0.683419,0.743476,0.404433); rgb(44pt)=(0.71841,0.741133,0.390476); rgb(45pt)=(0.752486,0.7384,0.376814); rgb(46pt)=(0.785843,0.735567,0.363271); rgb(47pt)=(0.818505,0.732733,0.34979); rgb(48pt)=(0.850657,0.7299,0.336029); rgb(49pt)=(0.882433,0.727433,0.3217); rgb(50pt)=(0.913933,0.725786,0.306276); rgb(51pt)=(0.944957,0.726114,0.288643); rgb(52pt)=(0.973895,0.731395,0.266648); rgb(53pt)=(0.993771,0.745457,0.240348); rgb(54pt)=(0.999043,0.765314,0.216414); rgb(55pt)=(0.995533,0.786057,0.196652); rgb(56pt)=(0.988,0.8066,0.179367); rgb(57pt)=(0.978857,0.827143,0.163314); rgb(58pt)=(0.9697,0.848138,0.147452); rgb(59pt)=(0.962586,0.870514,0.1309); rgb(60pt)=(0.958871,0.8949,0.113243); rgb(61pt)=(0.959824,0.921833,0.0948381); rgb(62pt)=(0.9661,0.951443,0.0755333); rgb(63pt)=(0.9763,0.9831,0.0538)},
colorbar,
colorbar style={ylabel style={font= \color{white!15!black}}, ylabel={E (a.u.)}},
]
\addplot [forget plot] graphics [xmin=0.5, xmax=474.5, ymin=0.5, ymax=474.5] {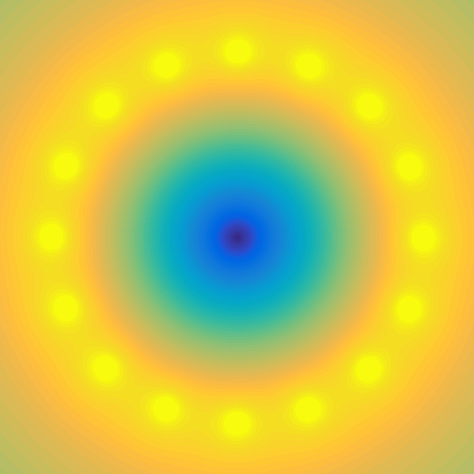};
\end{axis}

\begin{axis}[%
ticks=none,
width=1.4in,
height=1.4in,
at={(2.6in,1.533in)},
scale only axis,
point meta min=0,
point meta max=1,
axis on top,
xmin=0.5,
xmax=474.5,
xtick={100,200,300,400},
xticklabels={\empty},
y dir=reverse,
ymin=0.5,
ymax=474.5,
ytick={100,200,300,400},
yticklabels={\empty},
axis background/.style={fill=white},
title style={font=\bfseries},
title={$\text{$|$E}_\text{z}^{\text{inc}}\text{$|$}$},
colormap={mymap}{[1pt] rgb(0pt)=(0.2081,0.1663,0.5292); rgb(1pt)=(0.211624,0.189781,0.577676); rgb(2pt)=(0.212252,0.213771,0.626971); rgb(3pt)=(0.2081,0.2386,0.677086); rgb(4pt)=(0.195905,0.264457,0.7279); rgb(5pt)=(0.170729,0.291938,0.779248); rgb(6pt)=(0.125271,0.324243,0.830271); rgb(7pt)=(0.0591333,0.359833,0.868333); rgb(8pt)=(0.0116952,0.38751,0.881957); rgb(9pt)=(0.00595714,0.408614,0.882843); rgb(10pt)=(0.0165143,0.4266,0.878633); rgb(11pt)=(0.0328524,0.443043,0.871957); rgb(12pt)=(0.0498143,0.458571,0.864057); rgb(13pt)=(0.0629333,0.47369,0.855438); rgb(14pt)=(0.0722667,0.488667,0.8467); rgb(15pt)=(0.0779429,0.503986,0.838371); rgb(16pt)=(0.0793476,0.520024,0.831181); rgb(17pt)=(0.0749429,0.537543,0.826271); rgb(18pt)=(0.0640571,0.556986,0.823957); rgb(19pt)=(0.0487714,0.577224,0.822829); rgb(20pt)=(0.0343429,0.596581,0.819852); rgb(21pt)=(0.0265,0.6137,0.8135); rgb(22pt)=(0.0238905,0.628662,0.803762); rgb(23pt)=(0.0230905,0.641786,0.791267); rgb(24pt)=(0.0227714,0.653486,0.776757); rgb(25pt)=(0.0266619,0.664195,0.760719); rgb(26pt)=(0.0383714,0.674271,0.743552); rgb(27pt)=(0.0589714,0.683757,0.725386); rgb(28pt)=(0.0843,0.692833,0.706167); rgb(29pt)=(0.113295,0.7015,0.685857); rgb(30pt)=(0.145271,0.709757,0.664629); rgb(31pt)=(0.180133,0.717657,0.642433); rgb(32pt)=(0.217829,0.725043,0.619262); rgb(33pt)=(0.258643,0.731714,0.595429); rgb(34pt)=(0.302171,0.737605,0.571186); rgb(35pt)=(0.348167,0.742433,0.547267); rgb(36pt)=(0.395257,0.7459,0.524443); rgb(37pt)=(0.44201,0.748081,0.503314); rgb(38pt)=(0.487124,0.749062,0.483976); rgb(39pt)=(0.530029,0.749114,0.466114); rgb(40pt)=(0.570857,0.748519,0.44939); rgb(41pt)=(0.609852,0.747314,0.433686); rgb(42pt)=(0.6473,0.7456,0.4188); rgb(43pt)=(0.683419,0.743476,0.404433); rgb(44pt)=(0.71841,0.741133,0.390476); rgb(45pt)=(0.752486,0.7384,0.376814); rgb(46pt)=(0.785843,0.735567,0.363271); rgb(47pt)=(0.818505,0.732733,0.34979); rgb(48pt)=(0.850657,0.7299,0.336029); rgb(49pt)=(0.882433,0.727433,0.3217); rgb(50pt)=(0.913933,0.725786,0.306276); rgb(51pt)=(0.944957,0.726114,0.288643); rgb(52pt)=(0.973895,0.731395,0.266648); rgb(53pt)=(0.993771,0.745457,0.240348); rgb(54pt)=(0.999043,0.765314,0.216414); rgb(55pt)=(0.995533,0.786057,0.196652); rgb(56pt)=(0.988,0.8066,0.179367); rgb(57pt)=(0.978857,0.827143,0.163314); rgb(58pt)=(0.9697,0.848138,0.147452); rgb(59pt)=(0.962586,0.870514,0.1309); rgb(60pt)=(0.958871,0.8949,0.113243); rgb(61pt)=(0.959824,0.921833,0.0948381); rgb(62pt)=(0.9661,0.951443,0.0755333); rgb(63pt)=(0.9763,0.9831,0.0538)},
colorbar,
colorbar style={ylabel style={font= \color{white!15!black}}, ylabel={E (a.u.)}},
]
\addplot [forget plot] graphics [xmin=0.5, xmax=474.5, ymin=0.5, ymax=474.5] {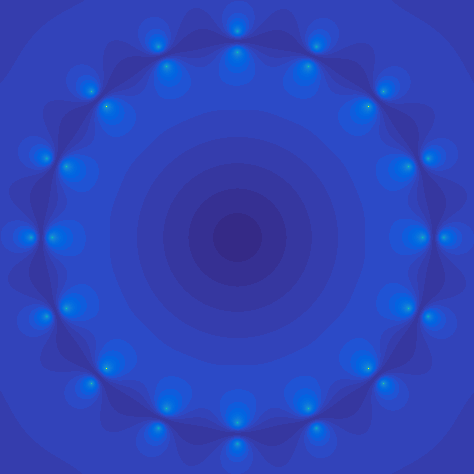};
\end{axis}
\end{tikzpicture}%
\caption{The top row shows the setup of the line sources (left) and the resulting incident electric field amplitude (right). The bottom row shows the same for the numerical implementation of the PEC with mirror sources. The same current is running through the line sources in both cases.}
\label{fig:rfshield}
\end{figure}
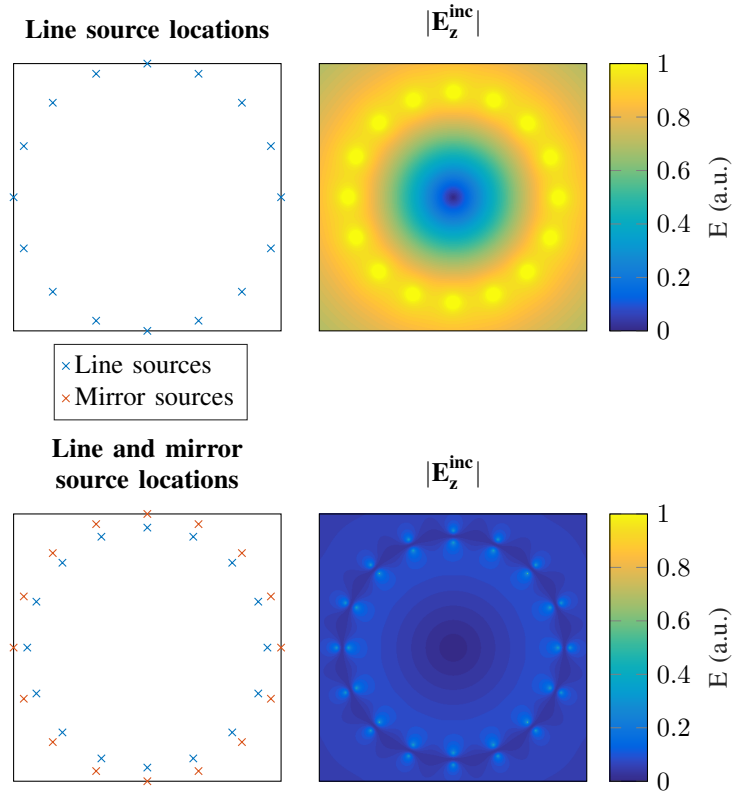

To avoid a severe increase in computation time of the CSI-EPT algorithm we implemented the RF-shield numerically by using mirror currents. At the location of the PEC the tangential electric field should be zero. To accomplish this for the circular RF-shield we assume that it is infinitely long in the $z$-direction. From there we follow \cite{jin1998electromagnetic}, where an improved placement of the mirror currents for a first order approximation of circular planes is given as
\begin{equation}
    d = \frac{R_{PEC}^2}{R_{S}},
\end{equation}
where $d$ is the distance from mirror source to the center of the birdcage coil, $R_{PEC}$ is the radius of the RF-shield and $R_{S}$ is the distance from the center of the RF-shield to the source. The effect of the mirror currents is shown in Figure \ref{fig:rfshield}.

From Figure \ref{fig:rfshield} it can be seen that the magnitude of the electric field significantly decreases for equal currents running through the line sources. Further, it can be seen that the electric field goes to zero at the location of the RF-shield. Including the RF-shield, even with a numerical approximation, results in a more realistic model of the incident RF fields which helps with the practical implementation of the CSI-EPT method.

\section{Methods}
\begin{figure}
%
%
\begin{tikzpicture}

\begin{axis}[%
ticks=none,
width=1.3in,
height=1.3in,
at={(1in,0.559in)},
scale only axis,
point meta min=-0,
point meta max=0.999999940395355,
axis on top,
xmin=0.5,
xmax=79.5,
xtick={20,40,60},
xticklabels={\empty},
y dir=reverse,
ymin=0.5,
ymax=79.5,
ytick={10,20,30,40,50,60,70},
yticklabels={\empty},
axis background/.style={fill=white},
title style={font=\bfseries},
title={Original conductivity},
legend style={legend cell align=left, align=left, draw=white!15!black},
colormap={mymap}{[1pt] rgb(0pt)=(0.2422,0.1504,0.6603); rgb(1pt)=(0.25039,0.164995,0.707614); rgb(2pt)=(0.257771,0.181781,0.751138); rgb(3pt)=(0.264729,0.197757,0.795214); rgb(4pt)=(0.270648,0.214676,0.836371); rgb(5pt)=(0.275114,0.234238,0.870986); rgb(6pt)=(0.2783,0.255871,0.899071); rgb(7pt)=(0.280333,0.278233,0.9221); rgb(8pt)=(0.281338,0.300595,0.941376); rgb(9pt)=(0.281014,0.322757,0.957886); rgb(10pt)=(0.279467,0.344671,0.971676); rgb(11pt)=(0.275971,0.366681,0.982905); rgb(12pt)=(0.269914,0.3892,0.9906); rgb(13pt)=(0.260243,0.412329,0.995157); rgb(14pt)=(0.244033,0.435833,0.998833); rgb(15pt)=(0.220643,0.460257,0.997286); rgb(16pt)=(0.196333,0.484719,0.989152); rgb(17pt)=(0.183405,0.507371,0.979795); rgb(18pt)=(0.178643,0.528857,0.968157); rgb(19pt)=(0.176438,0.549905,0.952019); rgb(20pt)=(0.168743,0.570262,0.935871); rgb(21pt)=(0.154,0.5902,0.9218); rgb(22pt)=(0.146029,0.609119,0.907857); rgb(23pt)=(0.138024,0.627629,0.89729); rgb(24pt)=(0.124814,0.645929,0.888343); rgb(25pt)=(0.111252,0.6635,0.876314); rgb(26pt)=(0.0952095,0.679829,0.859781); rgb(27pt)=(0.0688714,0.694771,0.839357); rgb(28pt)=(0.0296667,0.708167,0.816333); rgb(29pt)=(0.00357143,0.720267,0.7917); rgb(30pt)=(0.00665714,0.731214,0.766014); rgb(31pt)=(0.0433286,0.741095,0.73941); rgb(32pt)=(0.0963952,0.75,0.712038); rgb(33pt)=(0.140771,0.7584,0.684157); rgb(34pt)=(0.1717,0.766962,0.655443); rgb(35pt)=(0.193767,0.775767,0.6251); rgb(36pt)=(0.216086,0.7843,0.5923); rgb(37pt)=(0.246957,0.791795,0.556743); rgb(38pt)=(0.290614,0.79729,0.518829); rgb(39pt)=(0.340643,0.8008,0.478857); rgb(40pt)=(0.3909,0.802871,0.435448); rgb(41pt)=(0.445629,0.802419,0.390919); rgb(42pt)=(0.5044,0.7993,0.348); rgb(43pt)=(0.561562,0.794233,0.304481); rgb(44pt)=(0.617395,0.787619,0.261238); rgb(45pt)=(0.671986,0.779271,0.2227); rgb(46pt)=(0.7242,0.769843,0.191029); rgb(47pt)=(0.773833,0.759805,0.16461); rgb(48pt)=(0.820314,0.749814,0.153529); rgb(49pt)=(0.863433,0.7406,0.159633); rgb(50pt)=(0.903543,0.733029,0.177414); rgb(51pt)=(0.939257,0.728786,0.209957); rgb(52pt)=(0.972757,0.729771,0.239443); rgb(53pt)=(0.995648,0.743371,0.237148); rgb(54pt)=(0.996986,0.765857,0.219943); rgb(55pt)=(0.995205,0.789252,0.202762); rgb(56pt)=(0.9892,0.813567,0.188533); rgb(57pt)=(0.978629,0.838629,0.176557); rgb(58pt)=(0.967648,0.8639,0.16429); rgb(59pt)=(0.96101,0.889019,0.153676); rgb(60pt)=(0.959671,0.913457,0.142257); rgb(61pt)=(0.962795,0.937338,0.12651); rgb(62pt)=(0.969114,0.960629,0.106362); rgb(63pt)=(0.9769,0.9839,0.0805)},
colorbar horizontal,
colorbar style={ylabel style={font=\color{white!15!black}}, xlabel={$\sigma\text{ (S/m)}$}}
]
\addplot [forget plot] graphics [xmin=0.5, xmax=79.5, ymin=0.5, ymax=79.5] {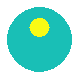};
\end{axis}

\begin{axis}[%
ticks=none,
width=1.3in,
height=1.3in,
at={(2.8in,0.559in)},
scale only axis,
point meta min=0,
point meta max=80,
axis on top,
xmin=0.5,
xmax=79.5,
xtick={20,40,60},
xticklabels={\empty},
y dir=reverse,
ymin=0.5,
ymax=79.5,
ytick={10,20,30,40,50,60,70},
yticklabels={\empty},
axis background/.style={fill=white},
title style={font=\bfseries},
title={Original permittivity},
legend style={legend cell align=left, align=left, draw=white!15!black},
colormap={mymap}{[1pt] rgb(0pt)=(0.2422,0.1504,0.6603); rgb(1pt)=(0.25039,0.164995,0.707614); rgb(2pt)=(0.257771,0.181781,0.751138); rgb(3pt)=(0.264729,0.197757,0.795214); rgb(4pt)=(0.270648,0.214676,0.836371); rgb(5pt)=(0.275114,0.234238,0.870986); rgb(6pt)=(0.2783,0.255871,0.899071); rgb(7pt)=(0.280333,0.278233,0.9221); rgb(8pt)=(0.281338,0.300595,0.941376); rgb(9pt)=(0.281014,0.322757,0.957886); rgb(10pt)=(0.279467,0.344671,0.971676); rgb(11pt)=(0.275971,0.366681,0.982905); rgb(12pt)=(0.269914,0.3892,0.9906); rgb(13pt)=(0.260243,0.412329,0.995157); rgb(14pt)=(0.244033,0.435833,0.998833); rgb(15pt)=(0.220643,0.460257,0.997286); rgb(16pt)=(0.196333,0.484719,0.989152); rgb(17pt)=(0.183405,0.507371,0.979795); rgb(18pt)=(0.178643,0.528857,0.968157); rgb(19pt)=(0.176438,0.549905,0.952019); rgb(20pt)=(0.168743,0.570262,0.935871); rgb(21pt)=(0.154,0.5902,0.9218); rgb(22pt)=(0.146029,0.609119,0.907857); rgb(23pt)=(0.138024,0.627629,0.89729); rgb(24pt)=(0.124814,0.645929,0.888343); rgb(25pt)=(0.111252,0.6635,0.876314); rgb(26pt)=(0.0952095,0.679829,0.859781); rgb(27pt)=(0.0688714,0.694771,0.839357); rgb(28pt)=(0.0296667,0.708167,0.816333); rgb(29pt)=(0.00357143,0.720267,0.7917); rgb(30pt)=(0.00665714,0.731214,0.766014); rgb(31pt)=(0.0433286,0.741095,0.73941); rgb(32pt)=(0.0963952,0.75,0.712038); rgb(33pt)=(0.140771,0.7584,0.684157); rgb(34pt)=(0.1717,0.766962,0.655443); rgb(35pt)=(0.193767,0.775767,0.6251); rgb(36pt)=(0.216086,0.7843,0.5923); rgb(37pt)=(0.246957,0.791795,0.556743); rgb(38pt)=(0.290614,0.79729,0.518829); rgb(39pt)=(0.340643,0.8008,0.478857); rgb(40pt)=(0.3909,0.802871,0.435448); rgb(41pt)=(0.445629,0.802419,0.390919); rgb(42pt)=(0.5044,0.7993,0.348); rgb(43pt)=(0.561562,0.794233,0.304481); rgb(44pt)=(0.617395,0.787619,0.261238); rgb(45pt)=(0.671986,0.779271,0.2227); rgb(46pt)=(0.7242,0.769843,0.191029); rgb(47pt)=(0.773833,0.759805,0.16461); rgb(48pt)=(0.820314,0.749814,0.153529); rgb(49pt)=(0.863433,0.7406,0.159633); rgb(50pt)=(0.903543,0.733029,0.177414); rgb(51pt)=(0.939257,0.728786,0.209957); rgb(52pt)=(0.972757,0.729771,0.239443); rgb(53pt)=(0.995648,0.743371,0.237148); rgb(54pt)=(0.996986,0.765857,0.219943); rgb(55pt)=(0.995205,0.789252,0.202762); rgb(56pt)=(0.9892,0.813567,0.188533); rgb(57pt)=(0.978629,0.838629,0.176557); rgb(58pt)=(0.967648,0.8639,0.16429); rgb(59pt)=(0.96101,0.889019,0.153676); rgb(60pt)=(0.959671,0.913457,0.142257); rgb(61pt)=(0.962795,0.937338,0.12651); rgb(62pt)=(0.969114,0.960629,0.106362); rgb(63pt)=(0.9769,0.9839,0.0805)},
colorbar horizontal,
colorbar style={ylabel style={font=\color{white!15!black}}, xlabel={$\epsilon{}_\text{r}$}}
]
\addplot [forget plot] graphics [xmin=0.5, xmax=79.5, ymin=0.5, ymax=79.5] {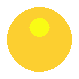};
\end{axis}

\begin{axis}[%
ticks=none,
width=1.3in,
height=1.3in,
at={(1in,-1.9in)},
scale only axis,
y dir=reverse,
ymin=0.5,
ymax=79.5,
ytick={10,20,30,40,50,60,70},
yticklabels={\empty},
xmin=0.5,
xmax=79.5,
xtick={20,40,60},
xticklabels={\empty},
axis background/.style={fill=white},
title style={font=\bfseries},
title={Geometry}
]
\draw[black, thick](0.65in,-0.65in) circle (0.52in);
\draw[black, thick](0.65in,-0.45in) circle (0.17in);
\node[black] at (0.65in,-0.45in) {Tube};
\node[black] at (0.65in,-0.75in) {Background};
\end{axis}

\node[] at (3.4in,-1.1in){
\scalebox{0.65}{
        \begin{tabular}{c|c|c|c}
            & \makecell{Conductivity \\(S/m)} & \makecell{Relative\\ Permittivity }& \makecell{Diameter \\(cm)}\\
            \hline
            \makecell{Tube\\ simulation} & 1 & 78 & 4\\
            \hline
            \makecell{Background\\ simulation} & 0.5 & 70& 12\\
        \end{tabular}}
    };

\end{tikzpicture}%
\caption{The top left shows the conductivity of the phantom and the top right shows the relative permittivity. The bottom left shows the the geometry of the phantom where the inner compartment will be referred to as the tube and the other compartment is referred to as the background. On the bottom right the table shows the conductivity and permittivity values that were used in simulations as well as the dimensions of the phantom.}
\label{fig:phantom}
\end{figure}
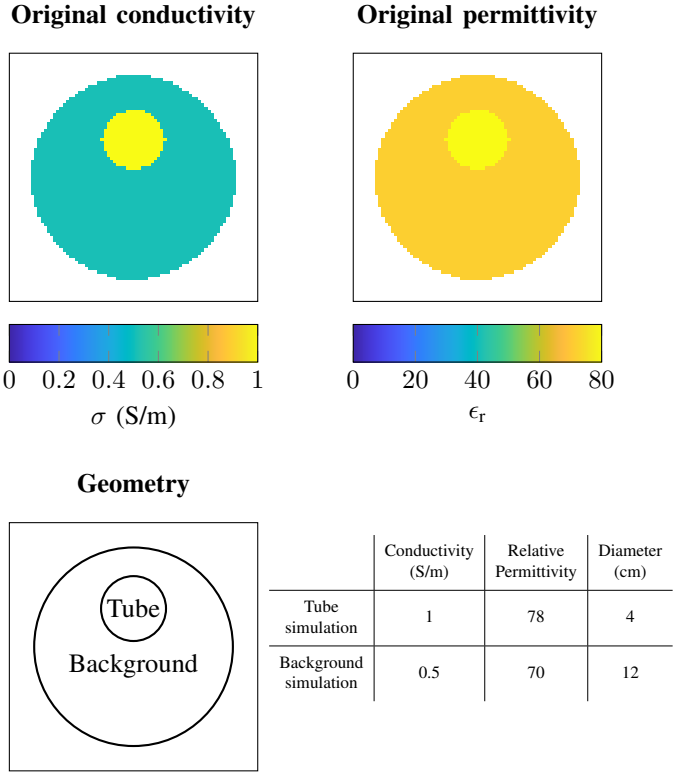

The performance of the improved updating scheme is tested first by use of 2D line source simulations in Matlab (MathWorks, Natick, Ma). The contrast that is used for these simulations is shown in Figure \ref{fig:phantom}. This contrast was chosen to be asymmetrical to render the TPA invalid. The incident electric and magnetic fields are computed using the setup that is shown in Figure \ref{fig:rfshield} with the RF-shield  included. Only the quadrature mode of the birdcage coil is used for reconstructions, linear modes are not included (i.e. $l = 1$).

In all the simulations the input is corrupted with white gaussian noise using realistic SNR values \cite{SNRthingy} for the corresponding field strength. The noise was added to the simulation by duplicating the $B_1$ maps and corrupting both the real and imaginary parts separately. From one pair of corrupted real and imaginary values the amplitude was extracted. The other pair of real and imaginary values was corrupted with a different noise set and used to construct the transceive phase. We follow this approach because the $B_1^+$ amplitude and transceive phase are acquired using two different measurements each with a corresponding noise set  \cite{Haacke1991}. Furthermore, in all the simulations total variation regularization was used during the minimization, as described in \cite{Balidemaj2015}. 

First, the effects of the transceive phase on the original CSI-EPT algorithm, using the TPA, and the newly proposed transceive phase corrected (TPC) CSI-EPT algorithm are investigated for 3T and 7T. After this, the effect the noise and the TPA have on the reconstruction of the conductivity at different field strengths is investigated. The SNR values that are used for these different field strengths are taken from \cite{SNRthingy}. 

To obtain a measure of the quality of the reconstructed electrical properties the mean absolute error is computed. This is performed within a region of the phantom as indicated by the black box in Figure \ref{fig:ErrorvsFS}. We define the mean absolute error as 
\begin{equation}
    Err_{\sigma} = \frac{1}{N}\sum\limits_N \frac{|\sigma_{true} - \sigma_{recon}|}{\sigma_{true}}\cdot 100\%,
\end{equation}
where $Err_{\sigma}$ defines the mean absolute error in the conductivity, $\sigma_{true}$ is the original conductivity of the phantom, $\sigma_{recon}$ is the reconstructed conductivity and $N$ is the number of points within the region indicated by the black box.

After these 2D line source simulations a 3D FDTD simulation package (Sim4Life, ZMT, Zurich, Switzerland) is used to investigate the effects of the 2D assumption that is made. For these simulations the same phantom as for the 2D line source simulations was constructed, the specifications are shown in Figure \ref{fig:phantom}. The transmit coil that is simulated is a 16 rung highpass birdcage coil with a diameter of 72cm and a rung length of 42cm.

Finally, a measurement using a 3T system (Philips, Igenia) is performed to show the potential of this method on in vivo data. Since there is no ground truth available with measurements, unlike the simulations, the CSI-EPT reconstructions are compared to the standard Helmholtz based EPT method, where a 7 point kernel is used \cite{Smandija17,VanLier2012}.

For the measurements a phantom was constructed with the same dimensions as the phantoms shown in Figure \ref{fig:phantom}. The phantom was agar based and NaCl was added to give the two compartments different conductivity values. In the tube 5.5 gr/L of NaCl was added this leading to a conductivity of 0.9 S/m at 21 $^\circ$C. For the background 2.5 gr/L was added resulting in a conductivity of 0.41 S/m at 21 $^\circ$C \cite{Stogryn}.

For the measurement an AFI sequence was used to obtain the $B_1^+$ amplitude \cite{Yarnykh2007}. The transceive phase was acquired with two spin echoes with opposite gradient polarity, this reduces the phase contribution due to the eddy currents \cite{VanLier2012}. The body coil was used for transmission and a head coil was used for reception.
A vendor specific algorithm (Philips, Constant Level of Appearance-CLEAR) was used to convert the receive phase measured with the head coil to the body coil. Using this algorithm it is as if the body coil was used for both transmitting and reception. The benefit is that using the head coil during reception significantly increases the SNR of the measurements. The sequence parameters that were used are noted in Table \ref{tab:seqparameters}.

\bgroup
\def\arraystretch{1.5}
\begin{table}
    \centering
    \caption{Scanner parameters for the AFI and SE sequence.}
    \begin{tabular}{c|c|c}
    Parameter   & AFI & SE \\
    \hline 
    FoV         & 200x200 mm$^2$     & 200x200 mm$^2$    \\
    \hline 
    Resolution  & 2.5x2.5x3mm$^3$    & 2.5x2.5x2.5mm$^3$    \\
    \hline 
    TR1         & 50ms               & 1000ms   \\
    \hline 
    TR2         & 250ms              & -    \\
    \hline 
    TE          & 2.7ms              &  5ms  \\
    \hline 
    Flip angle  & 65$^{\circ}$       & 90$^{\circ}$/180$^{\circ}$  \\
    \end{tabular} 
    \label{tab:seqparameters}
\end{table} 
\egroup

\section{Results}
\begin{figure*}
    \input{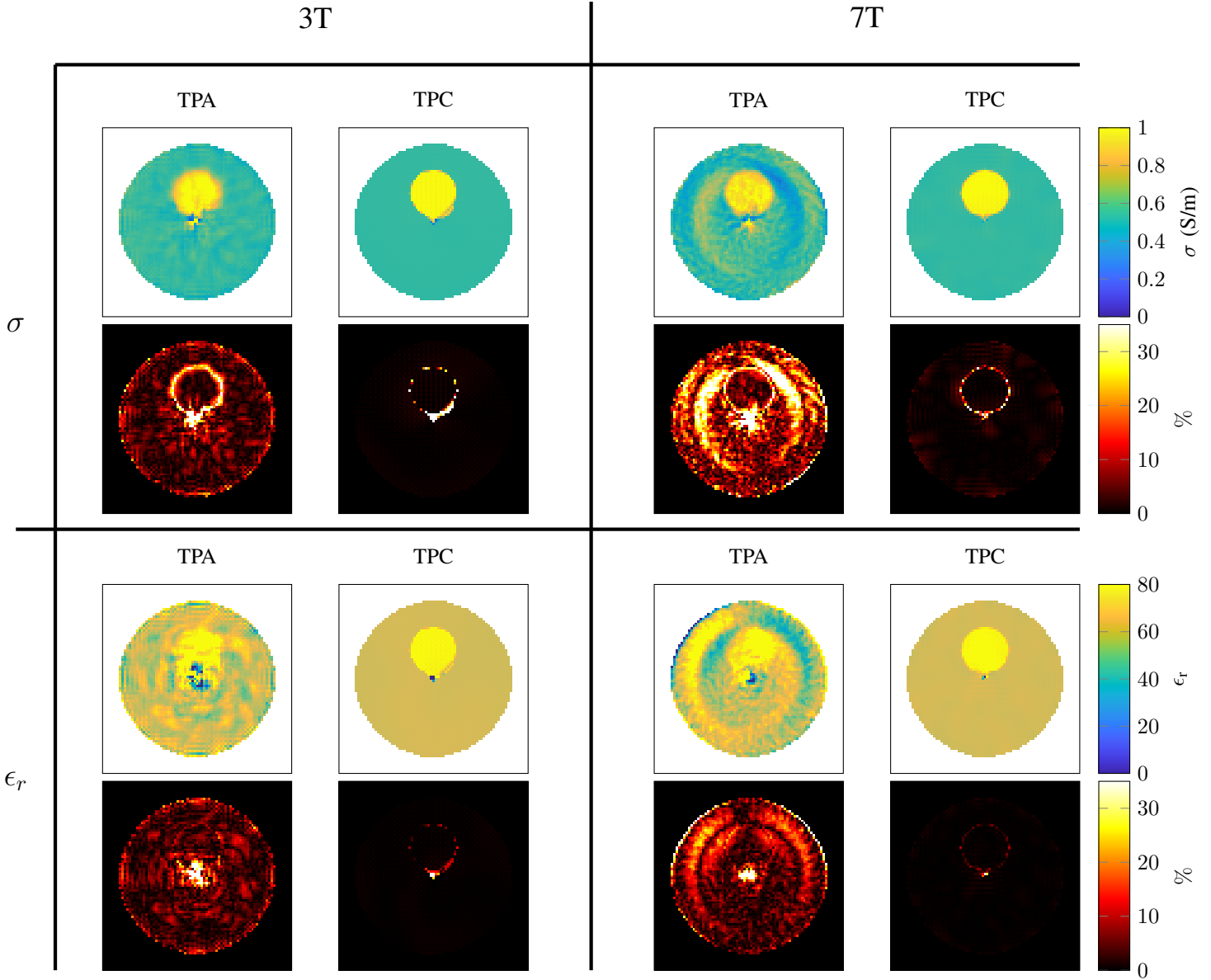}
    \caption{The top row shows the conductivity reconstructions with the corresponding error maps below them. The reconstructions were performed with the TPA and the newly proposed method,indicated with TPC, both at 3T and 7T. The third row shows the permittivity reconstructions with the corresponding error maps below them.}
    \label{fig:recons}
\end{figure*}

The results from the 2D line source simulations are shown in Figure \ref{fig:recons}. It can be observed that the error due to the TPA increases with increasing field strength. This is because of the increasing invalidity of the TPA at higher field strengths. Whereas for the newly proposed CSI-EPT algorithm this error is not present and the reconstruction improves due to the higher field strength with its inherently increased sensitivity. 

This is further illustrated in Figure \ref{fig:ErrorvsFS} where the effect of these two factors that are dependent on the field strength is shown. The mean absolute error of the reconstructions at different field strengths for both the standard CSI-EPT scheme and the newly proposed TPC CSI-EPT are shown. Furthermore, examples of the with noise corrupted simulated $B_1^+$ magnitude and transceive phase are shown.

Figure \ref{fig:outofmiddle} shows the conductivity reconstructions from the 3D FDTD simulations. To show the effect the 3D electromagnetic fields on the reconstructed contrast the input of the algorithm was not corrupted with noise. The top left reconstruction is taken from the center slice of the birdcage coil. The subsequent reconstructions are from slices 1cm  further out of the middle. All these reconstructions were computed separately as 2D slices and not as a 3D volume.

The bottom plot of Figure \ref{fig:outofmiddle} shows the conductivity value across the reconstructions for the point indicated with the red cross in the top left reconstruction. From this plot it can be seen that the conductivity value is underestimated. The underestimation increases for more peripheral locations along the cylindrical axis of the birdcage coil. This arises from the fact that the 2D EM field approximation becomes increasingly more invalid. Furthermore, the phantom is not 2D thus at the end of the phantom the RF field is not 2D E-polarized.

Figure \ref{fig:measurement} shows the reconstructed conductivity of MRI measured data for the proposed CSI-EPT method and the standard Helmholtz MR-EPT. For both methods a reconstruction was performed using a number of signal averages (NSA) of 2 and 10. For the NSA = 10 case the SNR is a factor $\sqrt{5}$ higher compared to the NSA = 2 case \cite{haacke1999magnetic}. The comparison between the two EPT methods was performed because there is no ground truth of the contrast available while this is the case with the simulations.

Since the newly proposed method also reconstructs the receive phase during the minimization process, it is possible to compare this reconstructed phase from the measurement with the 2D line source simulated one. This together with the comparison between the 2D simulated transceive phase and the measured transceive phase is shown in Figure \ref{fig:phases}.

Finally, in Table \ref{tab:MeanandSD} the mean ($\mu$) and standard deviation ($\sigma$) are given for the reconstructions in Figure \ref{fig:measurement}. These values are computed for the two different compartments of the phantom. It is known that standard Helmholtz MR-EPT is not able to reconstruct the boundaries properly \cite{Smandija17}, however the boundaries were included to compute the mean and standard deviation.

\begin{figure*}
\centering
\scalebox{0.39}{
    \input{figure4.tex}
}
    \caption{The top left shows an example of a simulated noisy $B_1^+$ amplitude map at 3T. The top middle figure shows the corresponding simulated noisy transceive phase. The top right figure shows where the mean absolute error was computed. The bottom figures show the mean absolute error in the conductivity and permittivity on a log scale versus the static magnetic field strength. Where the bottom left figure shows this relation when the TPA is used while the right shows this for the newly proposed TPC. The 3T and 7T reconstructions of these data points can be seen in Figure \ref{fig:recons}.}
    \label{fig:ErrorvsFS}
\vspace{0.5cm}
\scalebox{0.39}{
    \input{figure5.tex}
}
    \caption{The top left figure shows the reconstruction of the conductivity in the center of the birdcage coil for the 3D FDTD simulations. The seven subsequent figures are reconstructions each 1cm more out of the center slice of the birdcage coil. The bottom figure shows the value of the actual conductivity and the reconstructed value at the red cross marked in the top left figure.}
    \label{fig:outofmiddle}
\end{figure*}

\begin{figure}
%
%
\begin{tikzpicture}

\begin{axis}[%
width=1.35in,
height=1.35in,
at={(1.5in,1.8in)},
scale only axis,
point meta min=0,
point meta max=1,
axis on top,
xmin=0.5,
xmax=46.5,
xtick={\empty},
xticklabels={\empty},
y dir=reverse,
ymin=0.5,
ymax=46.5,
ytick={\empty},
yticklabels={\empty},
axis background/.style={fill=white},
title style={font=\bfseries, align=center},
title={  CSI-EPT \\   with NSA 10},
colormap={mymap}{ rgb=(0.2422,0.1504,0.6603); rgb=(0.25039,0.164995,0.707614); rgb=(0.257771,0.181781,0.751138); rgb=(0.264729,0.197757,0.795214); rgb=(0.270648,0.214676,0.836371); rgb=(0.275114,0.234238,0.870986); rgb=(0.2783,0.255871,0.899071); rgb=(0.280333,0.278233,0.9221); rgb=(0.281338,0.300595,0.941376); rgb=(0.281014,0.322757,0.957886); rgb=(0.279467,0.344671,0.971676); rgb=(0.275971,0.366681,0.982905); rgb=(0.269914,0.3892,0.9906); rgb=(0.260243,0.412329,0.995157); rgb=(0.244033,0.435833,0.998833); rgb=(0.220643,0.460257,0.997286); rgb=(0.196333,0.484719,0.989152); rgb=(0.183405,0.507371,0.979795); rgb=(0.178643,0.528857,0.968157); rgb=(0.176438,0.549905,0.952019); rgb=(0.168743,0.570262,0.935871); rgb=(0.154,0.5902,0.9218); rgb=(0.146029,0.609119,0.907857); rgb=(0.138024,0.627629,0.89729); rgb=(0.124814,0.645929,0.888343); rgb=(0.111252,0.6635,0.876314); rgb=(0.0952095,0.679829,0.859781); rgb=(0.0688714,0.694771,0.839357); rgb=(0.0296667,0.708167,0.816333); rgb=(0.00357143,0.720267,0.7917); rgb=(0.00665714,0.731214,0.766014); rgb=(0.0433286,0.741095,0.73941); rgb=(0.0963952,0.75,0.712038); rgb=(0.140771,0.7584,0.684157); rgb=(0.1717,0.766962,0.655443); rgb=(0.193767,0.775767,0.6251); rgb=(0.216086,0.7843,0.5923); rgb=(0.246957,0.791795,0.556743); rgb=(0.290614,0.79729,0.518829); rgb=(0.340643,0.8008,0.478857); rgb=(0.3909,0.802871,0.435448); rgb=(0.445629,0.802419,0.390919); rgb=(0.5044,0.7993,0.348); rgb=(0.561562,0.794233,0.304481); rgb=(0.617395,0.787619,0.261238); rgb=(0.671986,0.779271,0.2227); rgb=(0.7242,0.769843,0.191029); rgb=(0.773833,0.759805,0.16461); rgb=(0.820314,0.749814,0.153529); rgb=(0.863433,0.7406,0.159633); rgb=(0.903543,0.733029,0.177414); rgb=(0.939257,0.728786,0.209957); rgb=(0.972757,0.729771,0.239443); rgb=(0.995648,0.743371,0.237148); rgb=(0.996986,0.765857,0.219943); rgb=(0.995205,0.789252,0.202762); rgb=(0.9892,0.813567,0.188533); rgb=(0.978629,0.838629,0.176557); rgb=(0.967648,0.8639,0.16429); rgb=(0.96101,0.889019,0.153676); rgb=(0.959671,0.913457,0.142257); rgb=(0.962795,0.937338,0.12651); rgb=(0.969114,0.960629,0.106362); rgb=(0.9769,0.9839,0.0805);rgb=(0.9769,0.9839,0.0805)},
colorbar,
colorbar style={ylabel style={font=\color{white!15!black}}, ylabel={ $\sigma\text{ (S/m)}$},ytick={0,0.2,...,1},yticklabel style={
 align=left,
        }},
colorbar/width = 4mm
]
\addplot [forget plot] graphics [xmin=0.5, xmax=46.5, ymin=0.5, ymax=46.5] {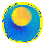};
\end{axis}

\begin{axis}[%
width=1.35in,
height=1.35in,
at={(0in,1.8in)},
scale only axis,
point meta min=0,
point meta max=1,
axis on top,
xmin=0.5,
xmax=46.5,
xtick={\empty},
xticklabels={\empty},
y dir=reverse,
ymin=0.5,
ymax=46.5,
ytick={\empty},
yticklabels={\empty},
axis background/.style={fill=white},
title style={font=\bfseries, align=center},
title={  CSI-EPT \\   with NSA 2}
]
\addplot [forget plot] graphics [xmin=0.5, xmax=46.5, ymin=0.5, ymax=46.5] {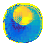};
\end{axis}

\begin{axis}[%
width=1.35in,
height=1.35in,
at={(1.5in,0in)},
scale only axis,
point meta min=0,
point meta max=1,
axis on top,
xmin=0.5,
xmax=46.5,
xtick={\empty},
xticklabels={\empty},
y dir=reverse,
ymin=0.5,
ymax=46.5,
ytick={\empty},
yticklabels={\empty},
axis background/.style={fill=white},
title style={font=\bfseries, align=center},
title={  Standard MR-EPT\\   with NSA 10},
colormap={mymap}{ rgb=(0.2422,0.1504,0.6603); rgb=(0.25039,0.164995,0.707614); rgb=(0.257771,0.181781,0.751138); rgb=(0.264729,0.197757,0.795214); rgb=(0.270648,0.214676,0.836371); rgb=(0.275114,0.234238,0.870986); rgb=(0.2783,0.255871,0.899071); rgb=(0.280333,0.278233,0.9221); rgb=(0.281338,0.300595,0.941376); rgb=(0.281014,0.322757,0.957886); rgb=(0.279467,0.344671,0.971676); rgb=(0.275971,0.366681,0.982905); rgb=(0.269914,0.3892,0.9906); rgb=(0.260243,0.412329,0.995157); rgb=(0.244033,0.435833,0.998833); rgb=(0.220643,0.460257,0.997286); rgb=(0.196333,0.484719,0.989152); rgb=(0.183405,0.507371,0.979795); rgb=(0.178643,0.528857,0.968157); rgb=(0.176438,0.549905,0.952019); rgb=(0.168743,0.570262,0.935871); rgb=(0.154,0.5902,0.9218); rgb=(0.146029,0.609119,0.907857); rgb=(0.138024,0.627629,0.89729); rgb=(0.124814,0.645929,0.888343); rgb=(0.111252,0.6635,0.876314); rgb=(0.0952095,0.679829,0.859781); rgb=(0.0688714,0.694771,0.839357); rgb=(0.0296667,0.708167,0.816333); rgb=(0.00357143,0.720267,0.7917); rgb=(0.00665714,0.731214,0.766014); rgb=(0.0433286,0.741095,0.73941); rgb=(0.0963952,0.75,0.712038); rgb=(0.140771,0.7584,0.684157); rgb=(0.1717,0.766962,0.655443); rgb=(0.193767,0.775767,0.6251); rgb=(0.216086,0.7843,0.5923); rgb=(0.246957,0.791795,0.556743); rgb=(0.290614,0.79729,0.518829); rgb=(0.340643,0.8008,0.478857); rgb=(0.3909,0.802871,0.435448); rgb=(0.445629,0.802419,0.390919); rgb=(0.5044,0.7993,0.348); rgb=(0.561562,0.794233,0.304481); rgb=(0.617395,0.787619,0.261238); rgb=(0.671986,0.779271,0.2227); rgb=(0.7242,0.769843,0.191029); rgb=(0.773833,0.759805,0.16461); rgb=(0.820314,0.749814,0.153529); rgb=(0.863433,0.7406,0.159633); rgb=(0.903543,0.733029,0.177414); rgb=(0.939257,0.728786,0.209957); rgb=(0.972757,0.729771,0.239443); rgb=(0.995648,0.743371,0.237148); rgb=(0.996986,0.765857,0.219943); rgb=(0.995205,0.789252,0.202762); rgb=(0.9892,0.813567,0.188533); rgb=(0.978629,0.838629,0.176557); rgb=(0.967648,0.8639,0.16429); rgb=(0.96101,0.889019,0.153676); rgb=(0.959671,0.913457,0.142257); rgb=(0.962795,0.937338,0.12651); rgb=(0.969114,0.960629,0.106362); rgb=(0.9769,0.9839,0.0805);rgb=(0.9769,0.9839,0.0805)},
colorbar,
colorbar style={ylabel style={font=\color{white!15!black}}, ylabel={ $\sigma\text{ (S/m)}$},ytick={0,0.2,...,1},yticklabel style={
 align=left,
        }},
colorbar/width = 4mm
]
\addplot [forget plot] graphics [xmin=0.5, xmax=46.5, ymin=0.5, ymax=46.5] {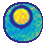};
\end{axis}

\begin{axis}[%
width=1.35in,
height=1.35in,
at={(0in,0in)},
scale only axis,
point meta min=0,
point meta max=1,
axis on top,
xmin=0.5,
xmax=46.5,
xtick={\empty},
xticklabels={\empty},
y dir=reverse,
ymin=0.5,
ymax=46.5,
ytick={\empty},
yticklabels={\empty},
axis background/.style={fill=white},
title style={font=\bfseries, align=center},
title={  Standard MR-EPT\\   with NSA 2}
]
\addplot [forget plot] graphics [xmin=0.5, xmax=46.5, ymin=0.5, ymax=46.5] {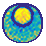};
\end{axis}
\end{tikzpicture}%
    \caption{The top left shows the conductivity reconstruction for the CSI-EPT reconstruction with an NSA of 2, the top right show the reconstruction for NSA = 10. The bottom two figures show the standard Helmholtz MR-EPT reconstruction. The left shows the NSA = 2 reconstruction while the right shows the NSA = 10 reconstruction.}
    \label{fig:measurement}
\end{figure}

\begin{figure}
%
%
\begin{tikzpicture}

\begin{axis}[%
width=1.35in,
height=1.35in,
at={(0in,0in)},
scale only axis,
point meta min=-5.27,
point meta max=-4.1,
axis on top,
xmin=0.5,
xmax=60.5,
xtick={\empty},
xticklabels={\empty},
y dir=reverse,
ymin=0.5,
ymax=60.5,
ytick={\empty},
yticklabels={\empty},
axis background/.style={fill=white},
title style={font=\bfseries},
title={Simulated $\phi_\text{+}$}
]
\addplot [forget plot] graphics [xmin=0.5, xmax=60.5, ymin=0.5, ymax=60.5] {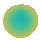};
\end{axis}

\begin{axis}[%
width=1.35in,
height=1.35in,
at={(1.5in,0in)},
scale only axis,
point meta min=0,
point meta max=1,
axis on top,
xmin=0.5,
xmax=80.5,
xtick={\empty},
xticklabels={\empty},
y dir=reverse,
ymin=0.5,
ymax=80.5,
ytick={\empty},
yticklabels={\empty},
axis background/.style={fill=white},
title style={font=\bfseries},
title={Reconstructed $\phi_\text{+}$},
colormap={mymap}{ rgb=(0.2422,0.1504,0.6603); rgb=(0.25039,0.164995,0.707614); rgb=(0.257771,0.181781,0.751138); rgb=(0.264729,0.197757,0.795214); rgb=(0.270648,0.214676,0.836371); rgb=(0.275114,0.234238,0.870986); rgb=(0.2783,0.255871,0.899071); rgb=(0.280333,0.278233,0.9221); rgb=(0.281338,0.300595,0.941376); rgb=(0.281014,0.322757,0.957886); rgb=(0.279467,0.344671,0.971676); rgb=(0.275971,0.366681,0.982905); rgb=(0.269914,0.3892,0.9906); rgb=(0.260243,0.412329,0.995157); rgb=(0.244033,0.435833,0.998833); rgb=(0.220643,0.460257,0.997286); rgb=(0.196333,0.484719,0.989152); rgb=(0.183405,0.507371,0.979795); rgb=(0.178643,0.528857,0.968157); rgb=(0.176438,0.549905,0.952019); rgb=(0.168743,0.570262,0.935871); rgb=(0.154,0.5902,0.9218); rgb=(0.146029,0.609119,0.907857); rgb=(0.138024,0.627629,0.89729); rgb=(0.124814,0.645929,0.888343); rgb=(0.111252,0.6635,0.876314); rgb=(0.0952095,0.679829,0.859781); rgb=(0.0688714,0.694771,0.839357); rgb=(0.0296667,0.708167,0.816333); rgb=(0.00357143,0.720267,0.7917); rgb=(0.00665714,0.731214,0.766014); rgb=(0.0433286,0.741095,0.73941); rgb=(0.0963952,0.75,0.712038); rgb=(0.140771,0.7584,0.684157); rgb=(0.1717,0.766962,0.655443); rgb=(0.193767,0.775767,0.6251); rgb=(0.216086,0.7843,0.5923); rgb=(0.246957,0.791795,0.556743); rgb=(0.290614,0.79729,0.518829); rgb=(0.340643,0.8008,0.478857); rgb=(0.3909,0.802871,0.435448); rgb=(0.445629,0.802419,0.390919); rgb=(0.5044,0.7993,0.348); rgb=(0.561562,0.794233,0.304481); rgb=(0.617395,0.787619,0.261238); rgb=(0.671986,0.779271,0.2227); rgb=(0.7242,0.769843,0.191029); rgb=(0.773833,0.759805,0.16461); rgb=(0.820314,0.749814,0.153529); rgb=(0.863433,0.7406,0.159633); rgb=(0.903543,0.733029,0.177414); rgb=(0.939257,0.728786,0.209957); rgb=(0.972757,0.729771,0.239443); rgb=(0.995648,0.743371,0.237148); rgb=(0.996986,0.765857,0.219943); rgb=(0.995205,0.789252,0.202762); rgb=(0.9892,0.813567,0.188533); rgb=(0.978629,0.838629,0.176557); rgb=(0.967648,0.8639,0.16429); rgb=(0.96101,0.889019,0.153676); rgb=(0.959671,0.913457,0.142257); rgb=(0.962795,0.937338,0.12651); rgb=(0.969114,0.960629,0.106362); rgb=(0.9769,0.9839,0.0805);rgb=(0.9769,0.9839,0.0805)},
colorbar,
colorbar/width = 4mm,
colorbar style={ylabel style={font=\color{white!15!black}}, ylabel={[a.u.]}}
]
\addplot [forget plot] graphics [xmin=0.5, xmax=80.5, ymin=0.5, ymax=80.5] {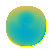};
\end{axis}

\begin{axis}[%
width=1.35in,
height=1.35in,
at={(0in,1.8in)},
scale only axis,
point meta min=-0.25,
point meta max=0.45,
axis on top,
xmin=0.5,
xmax=60.5,
xtick={\empty},
xticklabels={\empty},
y dir=reverse,
ymin=0.5,
ymax=60.5,
ytick={\empty},
yticklabels={\empty},
axis background/.style={fill=white},
title style={font=\bfseries},
title={Simulated $\phi_{\pm}$}
]
\addplot [forget plot] graphics [xmin=0.5, xmax=60.5, ymin=0.5, ymax=60.5] {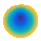};
\end{axis}

\begin{axis}[%
width=1.35in,
height=1.35in,
at={(1.5in,1.8in)},
scale only axis,
point meta min=0,
point meta max=1,
axis on top,
xmin=0.5,
xmax=80.5,
xtick={\empty},
xticklabels={\empty},
y dir=reverse,
ymin=0.5,
ymax=80.5,
ytick={\empty},
yticklabels={\empty},
axis background/.style={fill=white},
title style={font=\bfseries},
title={Measured $\phi_{\pm}$},
colormap={mymap}{ rgb=(0.2422,0.1504,0.6603); rgb=(0.25039,0.164995,0.707614); rgb=(0.257771,0.181781,0.751138); rgb=(0.264729,0.197757,0.795214); rgb=(0.270648,0.214676,0.836371); rgb=(0.275114,0.234238,0.870986); rgb=(0.2783,0.255871,0.899071); rgb=(0.280333,0.278233,0.9221); rgb=(0.281338,0.300595,0.941376); rgb=(0.281014,0.322757,0.957886); rgb=(0.279467,0.344671,0.971676); rgb=(0.275971,0.366681,0.982905); rgb=(0.269914,0.3892,0.9906); rgb=(0.260243,0.412329,0.995157); rgb=(0.244033,0.435833,0.998833); rgb=(0.220643,0.460257,0.997286); rgb=(0.196333,0.484719,0.989152); rgb=(0.183405,0.507371,0.979795); rgb=(0.178643,0.528857,0.968157); rgb=(0.176438,0.549905,0.952019); rgb=(0.168743,0.570262,0.935871); rgb=(0.154,0.5902,0.9218); rgb=(0.146029,0.609119,0.907857); rgb=(0.138024,0.627629,0.89729); rgb=(0.124814,0.645929,0.888343); rgb=(0.111252,0.6635,0.876314); rgb=(0.0952095,0.679829,0.859781); rgb=(0.0688714,0.694771,0.839357); rgb=(0.0296667,0.708167,0.816333); rgb=(0.00357143,0.720267,0.7917); rgb=(0.00665714,0.731214,0.766014); rgb=(0.0433286,0.741095,0.73941); rgb=(0.0963952,0.75,0.712038); rgb=(0.140771,0.7584,0.684157); rgb=(0.1717,0.766962,0.655443); rgb=(0.193767,0.775767,0.6251); rgb=(0.216086,0.7843,0.5923); rgb=(0.246957,0.791795,0.556743); rgb=(0.290614,0.79729,0.518829); rgb=(0.340643,0.8008,0.478857); rgb=(0.3909,0.802871,0.435448); rgb=(0.445629,0.802419,0.390919); rgb=(0.5044,0.7993,0.348); rgb=(0.561562,0.794233,0.304481); rgb=(0.617395,0.787619,0.261238); rgb=(0.671986,0.779271,0.2227); rgb=(0.7242,0.769843,0.191029); rgb=(0.773833,0.759805,0.16461); rgb=(0.820314,0.749814,0.153529); rgb=(0.863433,0.7406,0.159633); rgb=(0.903543,0.733029,0.177414); rgb=(0.939257,0.728786,0.209957); rgb=(0.972757,0.729771,0.239443); rgb=(0.995648,0.743371,0.237148); rgb=(0.996986,0.765857,0.219943); rgb=(0.995205,0.789252,0.202762); rgb=(0.9892,0.813567,0.188533); rgb=(0.978629,0.838629,0.176557); rgb=(0.967648,0.8639,0.16429); rgb=(0.96101,0.889019,0.153676); rgb=(0.959671,0.913457,0.142257); rgb=(0.962795,0.937338,0.12651); rgb=(0.969114,0.960629,0.106362); rgb=(0.9769,0.9839,0.0805);rgb=(0.9769,0.9839,0.0805)},
colorbar,
colorbar/width = 4mm,
colorbar style={ylabel style={font=\color{white!15!black}}, ylabel={[a.u.]}}
]
\addplot [forget plot] graphics [xmin=0.5, xmax=80.5, ymin=0.5, ymax=80.5] {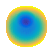};
\end{axis}
\end{tikzpicture}%
    \caption{The top left left shows the transceive phase from the 2D simulation. The top right shows the transceive phase measured with MRI. The bottom left shows the simulated transmit phase and the bottom right shows the transmit phase reconstructed by CSI-EPT from the measurement.}
    \label{fig:phases}
\end{figure}

\bgroup
\def\arraystretch{1.5}
\begin{table}
    \centering
    \caption{Mean and standard deviation (sd) of the MRI data reconstructions.}
    \begin{tabular}{c|c|c|c|c}
    \makecell{Reconstruction\\ used}   & \makecell{mean\\ inner tube} & \makecell{sd\\ inner tube} & \makecell{mean\\ background} & \makecell{sd\\ background} \\
    \hline 
    \makecell{CSI-EPT\\NSA = 2}         & 0.87     & 0.17 &0.23&0.12   \\
    \hline 
    \makecell{CSI-EPT\\NSA = 10}  & 0.87 & 0.05 & 0.23 & 0.09  \\
    \hline 
    \makecell{MR-EPT\\NSA = 2}        & 0.83 & 0.22 & 0.42  & 0.16 \\
    \hline 
    \makecell{MR-EPT\\NSA = 10}         & 0.81  & 0.22 & 0.37 & 0.13 \\
    \hline 
    \makecell{Reference\\Value} & 0.9 & - & 0.41 & -\\
    \end{tabular} 
    \label{tab:MeanandSD}
\end{table} 
\egroup

\section{Discussion}

In Figure \ref{fig:recons} it is shown that the reconstructions are improved significantly when the TPC CSI-EPT algorithm is used. Especially for the 7T reconstructions the transceive phase assumption error is predominant and the transceive phase correction significantly reduces the error.

The difference between the 3T and 7T TPC reconstructions arises from two factors, the first being the higher SNR that was simulated at the 7T field strength. The second factor is that the scattering currents are increased with increased frequency \cite{MaxwellEqs}. Therefore, the contrast has a higher electromagnetic imprint, making it easier to reconstruct.

In each of the eight reconstructions, in Figure \ref{fig:recons}, it can be noted that reconstruction in the center of the phantom has a larger error compared to the rest of the reconstruction. This is inherently due to the design of the birdcage coil, especially in quadrature and reverse quadrature mode. In these modes the electric fields constructed by each separate rung of the birdcage coil destructively interfere in the center of the coil creating an electric field that has an almost zero magnitude. This local minimum in the electric field, as can be seen in Figure \ref{fig:rfshield}, creates a singularity in the minimization and amplifies the noise in this region. To improve this either a different antenna setup could be chosen that creates a different electric field distribution or a dielectric pad could be used to move the local minimum in the electric field \cite{dielectricPadsWyger}.

Because the TPA is valid at lower field strengths Figure \ref{fig:ErrorvsFS} shows that for the lower field strengths the error in the reconstruction is dominated by the low SNR. For higher field strengths the error in the standard CSI-EPT algorithm increases because the TPA is no longer valid. In Figure \ref{fig:recons} these effects can also be observed. In the 3T reconstructions the error is predominantly due to the low SNR while at the 7T reconstruction more global over and underestimations of the conductivity and permittivity can be seen. When using the TPC CSI-EPT the transceive phase is no longer negatively affecting the reconstruction. Further, we observe that at the lower field strengths the conductivity reconstruction has a lower error compared to the permittivity reconstruction. The displacement current directly scales with the frequency \cite{MaxwellEqs}. Therefore, at higher static magnetic field strength the imprint of the permittivity on the contrast increases and is therefore reconstructed more easily.

Finally, from Figure 4 we like to point out the difference in the error in the reconstruction at the lower field strengths between the standard and TPC variants of the CSI-EPT algorithm. It is not exactly clear as to why this is the case. However, it could be that the total variation regularisation used performs better for the TPC minimization problem compared to the standard CSI-EPT. This should, however, be investigated further.

Figure \ref{fig:outofmiddle} shows how realistic three-dimensional electromagnetic fields affect the reconstruction. The center slice of the birdcage coil is the part that resembles a two-dimensional transverse magnetic (TM) polarized field. Slices outside of the center have larger $E_x$, $E_y$ and $B_z$ components that are not taken into account in this two-dimensional CSI-EPT algorithm. This is the cause of the underestimation of the conductivity as can be seen by the plot at the bottom of Figure \ref{fig:outofmiddle}, especially at the outermost slices of the phantom. At the boundary between the phantom and the air, the change in conductivity and permittivity creates three-dimensional scattered fields and the assumption that the field is TM polarized is no longer valid.

Possible solutions for the underestimation of the dielectric properties is to formulate the CSI-EPT algorithm for three-dimensional RF fields. However, this will significantly increase the computation time of the algorithm. Another solution could be to use a tube with a reference dielectric during scanning. Then the results can be scaled until the correct reference value is found. 

From the reconstruction of the measured data in Figure \ref{fig:measurement} it is clear that the CSI-EPT reconstruction is more noise robust compared to the standard Helmholtz MR-EPT. Another striking feature is that for EPT a clear boundary error is present while for CSI-EPT this error is not present. This boundary error for derivative based EPT arises from the finite difference based derivation calculation and the assumption of piecewise continuous dielectric properties \cite{Smandija17}. However, for the CSI-EPT algorithm we observe that for the measured data the conductivity is underestimated and the outer edge of the phantom is overestimated as was expected from the 3D FDTD reconstruction. A remark on this reconstruction is that the location of the phantom inside the MR was not exactly the same as for the FDTD simulations. Therefore, the minimum in the magnitude of the electric field is slightly below the inner tube, whereas in the FDTD simulations it was inside the inner tube. 
The minimum in the magnitude of the electric field amplifies the noise at that location. Therefore, for the CSI-EPT reconstruction with NSA = 2, the standard deviation, seen in Table \ref{tab:MeanandSD}, of the inner tube is increased significantly. This can be solved by using a different antenna setup, using a dielectric pad or increase the NSA in order to increase the SNR. Another explanation for this could be that the minimization procedure is overfitting this part of the contrast. Finally, from Table \ref{tab:MeanandSD} it can be seen that the mean conductivity that is found by CSI-EPT is underestimating the actual conductivity value. However, the CSI-EPT reconstruction is more precise than the standard Helmholtz MR-EPT reconstruction. 

Finally, Figure \ref{fig:phases} shows a comparison between the simulated and measured transceive phase. While overall the phase distributions look comparable it should be noted that at the edges of the phantom the curvature of the measured transceive phase is larger than that of the simulated one. This manifests itself in the reconstruction from Figure \ref{fig:measurement} as a larger conductivity value at the boundary of the phantom. In addition, from the comparison of the transceive phase the curvature of the phase in the background material is smaller. This would also explain why the value for the conductivity is underestimated in this area.

From the TPC CSI-EPT we can also extract the reconstructed transmit phase. In Figure \ref{fig:phases} the reconstructed transmit phase is compared to the simulated transmit phase in the two-dimensional Matlab simulation. However, because the simulated and measured transceive phases are not equal the reconstructed conductivity map is not correct. This in turn affects the reconstructed transmit phase, which shows overall a similar distribution compared to the simulated transmit phase, but the curvature of the phase is smaller.

With the standard CSI-EPT algorithm, and standard Helmholtz EPT, there was a trade off to be made for the field strength to use when measuring with a standard birdcage setup. At lower field strengths the SNR is poor while at higher field strengths the TPA is not valid. Both phenomena introduce errors in the reconstruction. With the TPC CSI-EPT this trade off is no longer present since it is not affected by the invalidity of the TPA. Therefore, the increased SNR and inherent sensitivity of 7T for a regular widely available standard quadrature setup can be exploited. 

The current limitation of the presented method is that for 3D RF fields the contrast is underestimated. To acquire better quantitative results a full fledged 3D CSI-EPT \cite{3DCSIEPT} would be required. The downside of using a 3D CSI-EPT reconstruction algorithm are the long computation times.

\section{Conclusion}
In this work the CSI-EPT algorithm was rewritten to use the transceive phase and $B_1^+$ amplitude as input instead of the complex $B_1^+$. Due to this reformulation the transceive phase assumption, which is not valid at high field strengths, is not necessary anymore. This allows for CSI-EPT to fully exploit the benefits of the higher static magnetic field strengths with a standard quadrature birdcage coil setup, these include a better SNR and increased RF field curvature due to the contrast. These benefits help the CSI-EPT algorithm to reconstruct the contrast more easily.

Further the effects of the three-dimensional nature of the electromagnetic fields on the two-dimensional CSI-EPT reconstruction have been explored. It was observed that the conductivity is underestimated, except for the edge of the contrast which is overestimated. Despite of this the first real MRI acquired data CSI-EPT reconstructions show significant improvement over the standard Helmholtz MR-EPT reconstructions. Moreover, we observed that CSI-EPT can reconstruct the boundaries between different dielectric properties and that it is robust with respect to realistic SNR values. The latter allows for either less signal averages during scanning, decreasing the acquisition time, or the contrast that is reconstructed to be more precise.



%





\ifCLASSOPTIONcaptionsoff
  \newpage
\fi

\end{document}